\documentclass[preprint,amsmath,amssymb,amsfonts]{revtex4}

\begin{document}

\begin{center}
{\bf Multiqubit Quantum Teleportation}
\end{center}

\begin{center}
Ming-Jing Zhao$^{1}$, {Zong-Guo Li$^{2}$},
{Xianqing Li-Jost$^{1}$} and {Shao-Ming Fei$^{1,3}$}\vspace{2ex}

\begin{minipage}{5in}

\small $~^{1}$ {Max-Planck-Institute for Mathematics in the
Sciences, 04103, Leipzig, Germany }

{\small $~^{2}$ College of Science, Tianjin University of
Technology, Tianjin, 300191, China}

{\small $~^{3}$ School of Mathematical Sciences, Capital Normal
University, Beijing 100048, China}

\end{minipage}
\end{center}

{E-mail: zhaomingjingde@126.com (Ming-Jing Zhao)}

{Abstract\\
We provide a class of six-qubit states for three-qubit perfect
teleportation. These states include the six-qubit cluster states as
a special class. We generalize this class of six-qubit states to
$2n$-qubit pure states for $n$-qubit teleportation, $n\geq 1$.
These states can be also used for $2n$ bit classical information
transmission in dense coding.}

PAC numbers: {03.65.Ud, 03.67.Ac}

\section{Introduction}

Quantum teleportation, employing classical communication and shared
resource of entanglement, allows to transmit an unknown quantum
state from a sender to a receiver that are spatially separated. Let
$|\phi\rangle$ be an arbitrary unknown pure state that is to be sent
from Alice to Bob, and $|\psi\rangle$ the entangled state shared by
Alice and Bob. To carry out teleportation Alice needs to perform
projective measurements on her two particles: one in the state
$|\phi\rangle$ and one part of the entangled state $|\psi\rangle$.
Learning the measurement results from Alice via the classical
communication channel, Bob applies a corresponding unitary
transformation on the other part of the entangled state
$|\psi\rangle$, so as to transform the state of this part to the
unknown state $|\phi\rangle$. In this scenario, Bennett {\it et.
al.} \cite{C. H. Bennett} first demonstrates the teleportation of an
arbitrary qubit state in terms of an entangled
Einstein-Podolsky-Rosen pair. Then the three-qubit GHZ state and a
class of W states are revealed to be the ideal resource for faithful
teleportation of one-qubit state \cite{A. Karlsson,P. Agrawal}. For
two-qubit teleportation, the tensor product of two Bell states
\cite{G. Rigolin}, the genuine four-qubit entangled state \cite{Y.
Yeo} and one five-qubit state in Ref. \cite{S. Muralidharan} are showed to
have the ability for faithful teleportation. Ref. \cite{J. Lee} has also proposed a scheme for two-qubit teleportation.
For three-qubit teleportation, Refs. \cite{S. Choudhury,X. Yin} have investigated the teleportation by one genuine entangled six-qubit state. Generally, Ref. \cite{P. X. Chen} provides the necessary and sufficient condition that the genuine $2n$-qubit entanglement channels must satisfy to teleport an arbitrary $n$-qubit state, and Ref. \cite{C. Y. Cheung} analyzes the criterion of multiqubit states for $n$-qubit teleportation. Then one kind of entangled $2n$-qubit states has been presented for $n$-qubit teleportation \cite{S. Muralidharan2010}. Besides
potential applications in quantum communication, quantum error
correction, and one way quantum computation, the cluster states are
also of importance in quantum teleportation and dense coding
\cite{cluster}.

In this paper, we propose a class of $2n$-qubit states for
$n$-qubit ($n\geq 1$) perfect teleportation, which are also the
ideal resource for transmission of $2n$ bit classical information in
dense coding. This class of states are not equivalent to the tensor
product of Bell states. For three-qubit and two-qubit cases, this
class of states includes the cluster states as a special case.

\section{three-qubit teleportation}

We first consider the quantum teleportation of an arbitrary unknown
three-qubit state $|\phi\rangle_{A^\prime}^{(3)}$.
Suppose Alice and Bob share a priori three pairs of qubits in the state
\begin{equation}\label{six-qubit resource}
\begin{array}{rcl}
|\xi\rangle_{A B}^{(3,3)} =
\frac{1}{2\sqrt{2}}\sum_{K=0}^7 |\overrightarrow{K}\rangle_{A}^{(3)}\otimes
|\overrightarrow{K}^\prime \rangle_{B}^{(3)},
\end{array}
\end{equation}
where $|\overrightarrow{K}\rangle^{(3)}$'s constitute an orthonormal basis,
\begin{equation*}
\begin{array}{rcl}
|\overrightarrow{0}\rangle_{A}^{(3)}&=& |00\rangle \otimes (\cos \theta_1 |0\rangle +\sin \theta_1 |1\rangle),\\
|\overrightarrow{1}\rangle_{A}^{(3)}&=&|00\rangle \otimes ( -\sin \theta_1 |0\rangle +\cos \theta_1 |1\rangle),\\
|\overrightarrow{2}\rangle_{A}^{(3)}&=& |01\rangle \otimes(\cos \theta_2 |0\rangle +\sin \theta_2 |1\rangle),\\
|\overrightarrow{3}\rangle_{A}^{(3)}&=& |01\rangle \otimes(\sin\theta_2 |0\rangle-\cos \theta_2 |1\rangle),\\
|\overrightarrow{4}\rangle_{A}^{(3)}&=& |10\rangle \otimes (\cos \theta_3 |0\rangle +\sin \theta_3 |1\rangle),\\
|\overrightarrow{5}\rangle_{A}^{(3)}&=&|10\rangle \otimes ( -\sin \theta_3 |0\rangle +\cos \theta_3 |1\rangle),\\
|\overrightarrow{6}\rangle_{A}^{(3)}&=& |11\rangle \otimes (\cos \theta_4 |0\rangle +\sin \theta_4 |1\rangle),\\
|\overrightarrow{7}\rangle_{A}^{(3)}&=&|11\rangle \otimes ( -\sin \theta_4 |0\rangle +\cos \theta_4 |1\rangle),
\end{array}
\end{equation*}
and $|\overrightarrow{K}^\prime \rangle^{(3)}$'s constitute another orthonormal basis,
\begin{equation*}
\begin{array}{rcl}
|\overrightarrow{0}^\prime\rangle_{B}^{(3)}&=& |00\rangle \otimes (\cos \theta^\prime_1 |0\rangle +\sin \theta^\prime_1 |1\rangle),\\
|\overrightarrow{1}^\prime\rangle_{B}^{(3)}&=&|00\rangle \otimes ( -\sin \theta^\prime_1 |0\rangle +\cos \theta^\prime_1 |1\rangle),\\
|\overrightarrow{2}^\prime\rangle_{B}^{(3)}&=& |01\rangle \otimes (\cos \theta^\prime_2 |0\rangle +\sin \theta^\prime_2 |1\rangle),\\
|\overrightarrow{3}^\prime\rangle_{B}^{(3)}&=&|01\rangle \otimes ( -\sin \theta^\prime_2 |0\rangle +\cos \theta^\prime_2 |1\rangle),\\
|\overrightarrow{4}^\prime\rangle_{B}^{(3)}&=& |10\rangle \otimes (\cos \theta^\prime_3 |0\rangle +\sin \theta^\prime_3 |1\rangle),\\
|\overrightarrow{5}^\prime\rangle_{B}^{(3)}&=&|10\rangle \otimes ( -\sin \theta^\prime_3 |0\rangle +\cos \theta^\prime_3 |1\rangle),\\
|\overrightarrow{6}^\prime\rangle_{B}^{(3)}&=& |11\rangle \otimes (\cos \theta^\prime_4 |0\rangle +\sin \theta^\prime_4 |1\rangle),\\
|\overrightarrow{7}^\prime\rangle_{B}^{(3)}&=&|11\rangle \otimes ( -\sin \theta^\prime_4 |0\rangle +\cos \theta^\prime_4 |1\rangle),
\end{array}
\end{equation*}
with $0\leq \theta_1, \theta_2, \theta_3, \theta^\prime_1,
\theta^\prime_2, \theta^\prime_3\leq \frac{\pi}{2}$. Here these orthonormal basis $\{|\overrightarrow{K} \rangle^{(3)}\}_{K=0}^7$ and $\{|\overrightarrow{K}^\prime \rangle^{(3)}\}_{K=0}^7$ can be viewed factually as the generalizations of the computational basis, and they will be shown to be able to give more resource for quantum teleportation. If we choose the basis for $A$ and $B$ as the three-qubit computational basis, then the resource for quantum teleportation in form of Eq. (\ref{six-qubit resource}) is naturally the tensor product of Bell states.

The arbitrary unknown three-qubit state $|\phi\rangle_{A^\prime}^{(3)}$ can be expressed
as
\begin{eqnarray*}
|\phi\rangle_{A^\prime}^{(3)}=\sum_{K=0}^7 a_K |\overrightarrow{K}^\prime\rangle_{A^\prime}^{(3)}
\end{eqnarray*}
with $\sum_{K=0}^7 |a_K|^2=1$. Noting that $|\xi\rangle_{AB}^{(3,3)}$ is a
maximally entangled state, we may construct the following basis of
$64$ orthonormal states:
\begin{equation}\label{pi}
\begin{array}{rcl}
|\Pi_{ijk}\rangle_{A^\prime A}^{(3,3)}=(\sigma^{(i)}\otimes\sigma^{(j)}\otimes\sigma^{(k)})_{A^\prime} |\Pi_{000}\rangle_{A^\prime
A}^{(3,3)},
\end{array}
\end{equation}
where $|\Pi_{000}\rangle_{A^\prime
A}^{(3,3)} =\frac{1}{2\sqrt{2}}\sum_{K=0}^7
|K^\prime\rangle_{A^\prime}^{(3)}\otimes |K
\rangle_{ A}^{(3)}$, $\sigma^{(0)}$ is the $2\times 2$ identity
matrix, $\sigma^{(1)}$, $\sigma^{(2)}$ and $\sigma^{(3)}$ are three
Pauli operators.
If Alice performs a complete projective measurement jointly on
$A^\prime A$ in the above basis
in Eq. (\ref{pi}) with the measurement outcome $ijk$, then Bob's sequences
of qubits will be in the state
$\sigma^{(i)}\otimes\sigma^{(j)}\otimes\sigma^{(k)}
|\phi\rangle_{B}^{(3)} $. Bob will always succeed in recovering
an exact replica of the original unknown state upon receiving 8 bits of classical
information about measurement results from Alice. Namely
\begin{eqnarray}
|\phi\rangle_{A^\prime}^{(3)}\otimes |\xi\rangle_{A B}^{(3,3)}
&=&\frac{1}{8}
\sum_{ijk}|\Pi_{ijk}\rangle_{A^\prime A}^{(3,3)}
(\sigma^{(i)}\otimes\sigma^{(j)}\otimes\sigma^{(k)})|\phi\rangle_{B}^{(3)},\\
^{(3,3)}_{A^\prime A}\langle \Pi_{ijk}|(|\phi\rangle_{A^\prime}^{(3)}\otimes |\xi\rangle_{A B}^{(3,3)})&=&\frac{1}{8}(\sigma^{(i)}\otimes\sigma^{(j)}\otimes\sigma^{(k)})|\phi\rangle_{B}^{(3)}.
\end{eqnarray}
These equations follow from the result given below, which also guarantees the success of the protocol.
\begin{eqnarray*}
^{(3,3)}_{A^\prime A}\langle \Pi_{000}|\xi\rangle_{A B}^{(3,3)}=\frac{1}{8} \sum_{J,K=0}^7 (^{(3)}_{A^\prime }\langle K^\prime|\otimes ^{(3)}_{A}\langle K|)(|J\rangle^{(3)}_{A }\otimes |J^\prime\rangle^{(3)}_{B})=\frac{1}{8}\sum_{K=0}^7 |K^\prime\rangle^{(3)}_{B } \times^{(3)}_{A^\prime }\langle K^\prime|.
\end{eqnarray*}

It can be verified that the reduced matrix
$\rho_{A_3B_3}=tr_{A_1A_2B_1 B_2}(|\xi\rangle_{A B}^{(3,3)}\langle \xi|)$ is not a pure state, as it is not rank one typically. This can be seen from its nonsingular submatrix spanned by $\{|00\rangle\langle00|,\ |00\rangle\langle11|\ |11\rangle\langle00|,\ |11\rangle\langle11|\}$: $[\cos^2(\theta_1-\theta^\prime_1)+\cos^2(\theta_2+\theta^\prime_2)+\cos^2(\theta_3-\theta^\prime_3)+
\cos^2(\theta_4-\theta^\prime_4)](|00\rangle\langle00|+|11\rangle\langle11|)+
[\cos^2(\theta_1-\theta^\prime_1)-\cos^2(\theta_2+\theta^\prime_2)+\cos^2(\theta_3-\theta^\prime_3)+
\cos^2(\theta_4-\theta^\prime_4)](|00\rangle\langle11|+|11\rangle\langle00|)$. Therefore,
$|\xi\rangle_{A B}^{(3)}$ is not equivalent to the
tensor product of Bell states, $\bigotimes_{i=1}^3
\frac{1}{\sqrt{2}}(|00\rangle+|11\rangle)_{A_iB_i}$. Furthermore, $\rho_{A_3B_3}$ is not maximally mixed state generally, so it is not equivalent to the genuine entangled six-qubit state in Refs. \cite{S. Choudhury,X. Yin}. When
$\theta_1=\theta_1^\prime$,
$\theta_2+\theta_2^\prime=\frac{\pi}{2}$,
$\theta_3=\theta_3^\prime$, $\theta_4=0$, and
$\theta_4^\prime=\frac{\pi}{2}$, the entangled state
$|\xi\rangle_{A B}^{(3,3)}$ becomes
\begin{equation*}
\begin{array}{rcl}
&&|\xi_0\rangle_{A B}^{(3,3)}\\
&=&\frac{1}{2\sqrt{2}}(|110111\rangle -|111110\rangle -
|101101\rangle +
|100100\rangle \\&&+ |000000\rangle
+|001001\rangle+|010011\rangle
+|011010\rangle)_{A_1A_2A_3 B_1B_2B_3}
\\ &=&
\frac{1}{2\sqrt{2}}((|00\rangle +|11\rangle)|0\rangle(|000\rangle +
|111\rangle)+(|00\rangle -|11\rangle)|1\rangle(|100\rangle +
|011\rangle))_{A_1B_1A_3 B_3 A_2B_2},
\end{array}
\end{equation*}
which is exactly the six-qubit cluster state \cite{cluster}.

Additionally, $|\xi\rangle_{A B}^{(3,3)}$ in Eq.
(\ref{six-qubit resource}) is also able to transmit $64$ bit
classical information. A dense coding scheme using $|\xi\rangle_{AB}^{(3,3)}$ is the following. Because $|\xi\rangle_{AB}^{(3,3)}$ is maximally entangled between
$A$ and $B$, $\{ (\sigma^{(i_1)}\otimes
\sigma^{(i_2)}\otimes \sigma^{(i_{3})} )_A|\xi\rangle_{AB}^{(3,3)}\}_{i_1,i_2,i_3=0}^3$ are $64$ maximally entangled states and
constitute an orthonormal basis for the $64$ dimensional Hilbert
space. The classical information sender Alice encodes her messages
by using operators $\sigma^{(i_1)}$, $\sigma^{(i_2)}$,
$\sigma^{(i_3)}$, and sends her qubits to Bob. Bob then
decodes the messages by performing a joint measurement on all six
qubits in the basis in Eq. (\ref{pi}). Here Alice  may encode her three qubits
locally and independently. While Bob is compelled to read the
messages from all the six-qubit together. This is different from a
straightforward extension of the original dense coding scheme in
terms of two Bell states, where Bob can measure his qubits
individually.

\section{multiqubit quantum teleportation}

Now we investigate the teleportation of an arbitrary $n$-qubit state
$|\phi\rangle_{A^\prime}^{(n)}$ by
extending the above ``generalized cluster-like" six-qubit states to
2n-qubit ones. A priori sequences of qubits shared by Alice and Bob are in the
state
\begin{equation}\label{2n qubit resource}
\begin{array}{rcl}
|\xi\rangle_{AB}^{(n,n)} =
\displaystyle\frac{1}{\sqrt{2^n}}\sum_{K=0}^{2^n-1}
|\overrightarrow{K}\rangle_{A}^{(n)}\otimes
|\overrightarrow{K^\prime} \rangle_{B}^{(n)},
\end{array}
\end{equation}
with $|\overrightarrow{K}\rangle^{(n)}$'s an orthonormal basis,
\begin{equation*}
\begin{array}{rcl}
|\overrightarrow{0}\rangle_{A}^{(n)}&=&
|\overrightarrow{0}\rangle_{A}^{(n-1)} \otimes (\cos
\theta_1 |0\rangle +\sin \theta_1 |1\rangle),\\
|\overrightarrow{1}\rangle_{A}^{(n)}&=&
|\overrightarrow{0}\rangle_{A}^{(n-1)} \otimes (
-\sin \theta_1 |0\rangle +\cos \theta_1 |1\rangle),\\
|\overrightarrow{2}\rangle_{A}^{(n)}&=&
|\overrightarrow{1}\rangle_{A}^{(n-1)} \otimes(\cos
\theta_2 |0\rangle +\sin \theta_2 |1\rangle),\\
|\overrightarrow{3}\rangle_{A}^{(n)}&=&
|\overrightarrow{1}\rangle_{A}^{(n-1)}
\otimes(\sin\theta_2 |0\rangle-\cos \theta_2 |1\rangle),\\
|\overrightarrow{4}\rangle_{A}^{(n)}&=&
|\overrightarrow{2}\rangle_{A}^{(n-1)} \otimes (\cos
\theta_3 |0\rangle +\sin \theta_3 |1\rangle),\\
|\overrightarrow{5}\rangle_{A}^{(n)}&=&|\overrightarrow{2}\rangle_{A}^{(n-1)} \otimes (
-\sin \theta_3 |0\rangle +\cos \theta_3 |1\rangle),\\
&\vdots&\\
|\overrightarrow{2^{n}-2}\rangle_{A}^{(n)}&=&
|\overrightarrow{2^{n-1}-1}\rangle_{A}^{(n-1)} \otimes
(\cos \theta_{2^{n-1}} |0\rangle +\sin \theta_{2^{n-1}}
|1\rangle),\\
|\overrightarrow{2^n-1}\rangle_{A}^{(n)}&=&|\overrightarrow{2^{n-1}-1}\rangle_{A}^{(n-1)}
\otimes ( -\sin \theta_{2^{n-1}} |0\rangle +\cos \theta_{2^{n-1}}
|1\rangle),
\end{array}
\end{equation*}
and $|\overrightarrow{K^\prime} \rangle^{(n)}$'s another orthonormal basis
\begin{equation*}
\begin{array}{rcl}
|\overrightarrow{0^\prime}\rangle_{B}^{(n)}&=&
|\overrightarrow{0^\prime}\rangle_{B}^{(n-1)} \otimes
(\cos \theta^\prime_1 |0\rangle +\sin \theta^\prime_1
|1\rangle),\\
|\overrightarrow{1^\prime}\rangle_{B}^{(n)}&=&|\overrightarrow{0^\prime}\rangle_{B}^{(n-1)}
\otimes ( -\sin \theta^\prime_1 |0\rangle +\cos \theta^\prime_1
|1\rangle),\\
|\overrightarrow{2^\prime}\rangle_{B}^{(n)}&=&
|\overrightarrow{1^\prime}\rangle_{B}^{(n-1)}
\otimes(\cos \theta^\prime_2 |0\rangle +\sin \theta^\prime_2
|1\rangle),\\
|\overrightarrow{3^\prime}\rangle_{B}^{(n)}&=&
|\overrightarrow{1^\prime}\rangle_{B}^{(n-1)} \otimes(
-\sin \theta^\prime_2 |0\rangle +\cos \theta^\prime_2
|1\rangle),\\
|\overrightarrow{4^\prime}\rangle_{B}^{(n)}&=&
|\overrightarrow{2^\prime}\rangle_{B}^{(n-1)} \otimes
(\cos \theta^\prime_3 |0\rangle +\sin \theta^\prime_3
|1\rangle),\\
|\overrightarrow{5^\prime}\rangle_{B}^{(n)}&=&|\overrightarrow{2^\prime}\rangle_{B}^{(n-1)}
\otimes ( -\sin \theta^\prime_3 |0\rangle +\cos \theta^\prime_3
|1\rangle),\\
&\vdots&\\
|\overrightarrow{(2^{n}-2)^\prime}\rangle_{B}^{(n)}&=&
|\overrightarrow{(2^{n-1}-1)^\prime}\rangle_{B}^{(n-1)}
\otimes (\cos \theta^\prime_{2^{n-1}} |0\rangle +\sin
\theta^\prime_{2^{n-1}} |1\rangle),\\
|\overrightarrow{(2^n-1)^\prime}\rangle_{B}^{(n)}&=&|\overrightarrow{(2^{n-1}-1)^\prime}\rangle_{B}^{(n-1)} \otimes ( -\sin \theta^\prime_{2^{n-1}} |0\rangle +\cos
\theta^\prime_{2^{n-1}} |1\rangle),
\end{array}
\end{equation*}
for $0\leq \theta_1, \theta^\prime_1, \cdots, \theta_{2^{n-1}},
\theta^\prime_{2^{n-1}}\leq \frac{\pi}{2}$.
$\{|\overrightarrow{K}\rangle_{A}^{(n-1)}
\}_{K=0}^{2^{n-1}-1}$ and $\{|\overrightarrow{K^\prime}\rangle_{B}^{(n-1)} \}_{K=0}^{2^{n-1}-1}$ are arbitrary orthonormal
basis of the $n-1$ qubit systems. Hence we may express the state to
be teleported $|\phi\rangle_{A}^{(n)}$ in the basis
$\{|\overrightarrow{K^\prime}\rangle_{B}^{(n)}
\}_{K=0}^{2^{n}-1}$,
\begin{eqnarray*}
|\phi\rangle_{A}^{(n)}=\sum_{K=0}^{2^n-1} a_K
|\overrightarrow{K^\prime}\rangle_{A^\prime}^{(n)}
\end{eqnarray*}
with $\sum_{K=0}^{2^n-1} |a_K|^2=1$. By virtue of the fact that
$|\xi\rangle_{AB}^{(n,n)}$ is a maximally
entangled state between $A$ and $B$, we may construct the following basis of $2^{2n}$
orthonormal states:
\begin{equation}\label{mul pi}
\begin{array}{rcl}
|\Pi_{i_1i_2\cdots i_n}\rangle_{A^\prime A}^{(n,n)}
=(\sigma^{(i_1)}\otimes\sigma^{(i_2)}\otimes\cdots
\otimes \sigma^{(i_n)})_{A^\prime} |\Pi_{00\cdots
0}\rangle_{A^\prime A}^{(n,n)},
\end{array}
\end{equation}
where
$$
|\Pi_{00\cdots 0}\rangle_{A^\prime A}^{(n,n)} =\frac{1}{\sqrt{2^n}}\sum_{K=0}^{2^n-1}
|\overrightarrow{K^\prime}\rangle_{A^\prime}^{(n)}\otimes |\overrightarrow{K}\rangle_{A}^{(n)},
$$
and $i_1, \cdots, i_n\in\{0,1,2,3\}$.

Now Alice performs a complete projective measurement jointly on
$A^\prime A$ in the
above basis in Eq. (\ref{mul pi}). If the measurement outcome is $i_1i_2\cdots i_n$,
then Bob's sequences of qubits will be in the
state $\sigma^{(i_1)}\otimes\sigma^{(i_2)}\otimes\cdots
\otimes \sigma^{(i_n)} |\phi\rangle_{B}^{(n)}$.
Bob can always recover an exact replica of the original unknown state, since
\begin{eqnarray}
|\phi\rangle_{A^\prime}^{(n)}\otimes
|\xi\rangle_{A B}^{(n,n)}
&=&\frac{1}{2^n} \sum_{i_1i_2\cdots i_n}|\Pi_{i_1i_2\cdots
i_n}\rangle_{A^\prime A}^{(n)} (\sigma^{(i_1)}\otimes\sigma^{(i_2)}\otimes\cdots
\otimes \sigma^{(i_n)})|\phi\rangle_{B}^{(n)},\\
^{(n,n)}_{A^\prime A}\langle \Pi_{00\cdots0}|(|\phi\rangle_{A^\prime}^{(n)}\otimes
|\xi\rangle_{A B}^{(n,n)})&=&\frac{1}{2^n}(\sigma^{(i_1)}\otimes\sigma^{(i_2)}\otimes\cdots
\otimes \sigma^{(i_n)})|\phi\rangle_{B}^{(n)}.
\end{eqnarray}
These equations follow from the result given below, which also guarantees the success of the protocol.
\begin{eqnarray*}
^{(n,n)}_{A^\prime A}\langle \Pi_{00\cdots0}|\xi\rangle_{A B}^{(n,n)}=\frac{1}{2^n} \sum_{J,K=0}^{2^{n}-1} (^{(n)}_{A^\prime }\langle K^\prime|\otimes ^{(n)}_{A}\langle K|)(|J\rangle^{(n)}_{A }\otimes |J^\prime\rangle^{(n)}_{B})=\frac{1}{2^n} \sum_{K=0}^{2^{n}-1} |K^\prime\rangle^{(n)}_{B } \times^{(n)}_{A^\prime }\langle K^\prime|.
\end{eqnarray*}
Therefore, $2n$-qubit state $|\xi\rangle_{A B}^{(n,n)}$ is an ideal resource for $n$-qubit teleportation.

As we know the tensor product of Bell state
$|\eta\rangle= \bigotimes_{i=1}^n |\psi^+\rangle_{A_iB_i}$
with $|\psi^+\rangle=\frac{1}{\sqrt{2}}(|00\rangle+|11\rangle)$, is the ideal resource for the $n$ qubit teleportation.
Here one can verify that the reduced matrix $\rho_{A_nB_n}=tr_{A_1A_2\cdots A_{n-1} B_1 B_2\cdots B_{n-1}}(|\xi\rangle_{A_1 A_2 \cdots A_n B_1 B_2 \cdots B_n}\langle \xi|)$ is not a pure state, therefore, $|\xi\rangle_{A B}^{(n,n)}$ is not equivalent to the tensor product of Bell states. This also shows $|\xi\rangle_{AB}^{(n,n)}$ is not equivalent to the entangled states used for $n$-qubit teleportation in Ref. \cite{S. Muralidharan2010}. Furthermore, when $n=2$, one can verify that four-qubit cluster state is a special case of $|\xi\rangle_{AB}^{(n,n)}$.

The state $|\xi\rangle_{AB}^{(n,n)}$ in Eq.
(\ref{2n qubit resource}) can be again used for dense coding and is
able to transmit $2n$ bit classical information. Since
$|\xi\rangle_{AB}^{(n,n)}$ is maximally
entangled between $A$ and $B$, $\{(\sigma^{(i_1)}\otimes \sigma^{(i_2)}\otimes
\cdots \otimes \sigma^{(i_{n})})_A |\xi\rangle_{AB}^{(n,n)}\}_{i_1,i_2,\cdots,i_n=0}^3$ are $2^{2n}$ maximally entangled states and
constitute an orthonormal basis for the $2^{2n}$ dimensional Hilbert
space. Alice can encode the message by using the operators
$\sigma^{(i_1)}$, $\sigma^{(i_2)}$, $\cdots$, and
$\sigma^{(i_n)}$. Upon receiving the qubits from Alice, Bob can
decode the message by performing a joint measurement in the
basis in Eq. (\ref{mul pi}).

\section{Conclusions}

We have presented a class of $2n$-qubit states which may serve
as ideal resources for perfect teleportation of $n$-qubit states
($n\geq 1$).  This class of states is also the ideal resource for
transmission of $2n$ bit classical information in dense coding. This
kind of states are not equivalent to the tensor product of Bell
states. For three-qubit and two-qubit cases, it is easy to see that
this class of states includes the cluster states as a special case.
Here we just study the teleportation and dense coding in terms of
these kinds of entangled states. It would be also interesting to
consider one-way quantum computing according to these ``generalized
cluster-like" states.

\end{document}